# Rotating topological edge solitons

**SERGEY K. IVANOV,**[1,*] **YAROSLAV V. KARTASHOV**[2]

[1]*ICFO-Institut de Ciencies Fotoniques, The Barcelona Institute of Science and Technology, 08860, Castelldefels (Barcelona), Spain*
[2]*Institute of Spectroscopy, Russian Academy of Sciences, 108840, Troitsk, Moscow, Russia*
*Corresponding author: sergey.ivanov@icfo.eu*

**We address the formation of topological edge solitons in rotating Su–Schrieffer–Heeger waveguide arrays. The linear spectrum of the non-rotating topological array is characterized by the presence of topological gap with two edge states residing in it. Rotation of the array significantly modifies the spectrum and may move these edge states out of the topological gap. Defocusing nonlinearity counteracts this tendency and shifts such modes back into topological gap, where they acquire structure of tails typical for topological edge states. We present rich bifurcation structure for rotating topological solitons and show that they can be stable. Rotation of the topologically trivial array, without edge states in its spectrum, also leads to the appearance of localized edge states, but in a trivial semi-infinite gap. Families of rotating edge solitons bifurcating from the trivial linear edge states exist too and sufficiently strong defocusing nonlinearity can also drive them into the topological gap, qualitatively modifying the structure of their tails.**

Topological insulation is a fundamental phenomenon encountered in several areas of physics. Topological insulators have been demonstrated in different platforms, starting from dimerized Su-Schrieffer-Heeger (SSH) lattices, to systems with broken time-reversal symmetry, time-periodic Floquet systems, various higher-order insulators, and many others. The majority of experiments with topological systems, conducted so far, were performed in linear regime (see reviews [1,2]). At the same time, in some physical systems, including photonic ones, the nonlinear effects arising upon increase of the amplitude of excitations may become strong enough to notably affect propagation dynamics of the topological edge states or to significantly alter topological phases (see review [3]).

One of the most striking manifestations of nonlinearity in topological insulators is the possibility of formation of topological edge solitons in them. They are unique states that exhibit topological protection and simultaneously feature a rich variety of shapes and interactions. In Optics, topological soliton-like edge states have been predicted [4] and observed [5] in Floquet systems [6-8]. Theory for envelopes of topological edge states has been developed for discrete [9,10] and continuous [11-13] helical waveguide arrays. Topological soliton metacrystals in a circular arrangement and their topological characterization in microresonators have been introduced in [14]. Dirac [15] and Bragg topological solitons [16,17] were predicted, as well as rich variety of corner solitons in second-order insulators [18,19]. Topological solitons have been encountered also in SSH arrays with straight waveguides [20-25], in topological trimer arrays [26], and in truncated photonic graphene [27].

In this Letter, using SSH structure with representative experimentally realistic parameters we study how the properties of topological edge solitons are affected by array rotation. We stress that previously the impact of rotation on the properties and stability of self-sustained states has been addressed only in topologically trivial structures [28-34]. Here we show that edge states appearing in topological gap in the spectrum of SSH array experience shift due to array rotation, and may even move outside this gap. Non-rotating trivial SSH arrays acquire localized edge states due to rotation, but in a semi-infinite gap. Edge solitons emanating from edge states in both topological and trivial rotating arrays under the action of *defocusing* nonlinearity undergo rich bifurcations. In some cases, if states leave the gap due to rotation, they can be shifted back into topological gap due to nonlinearity that is accompanied by qualitative modification of soliton tails. Thus, the interplay of rotation and nonlinearity leads to qualitative modification of the edge soliton properties and allows to induce topologically nontrivial phase.

We consider the propagation of light beams along the $z$ axis of a rotating (around the axis passing through its center) SSH waveguide array in a medium with defocusing cubic nonlinearity. In this case, the evolution of the dimensionless field amplitude $\psi$ is governed by the nonlinear Schrödinger equation:

$$i\frac{\partial \psi}{\partial z} = -\frac{1}{2}\left(\frac{\partial^2 \psi}{\partial x^2} + \frac{\partial^2 \psi}{\partial y^2}\right) + i\omega\left(x\frac{\partial \psi}{\partial y} - y\frac{\partial \psi}{\partial x}\right) - \mathcal{R}(x,y)\psi + |\psi|^2\psi$$

(1)

where $x = [X\cos(\omega z) + Y\sin(\omega z)]/w_0$ and $y = [Y\cos(\omega z) - X\sin(\omega z)]/w_0$ are the transverse coordinates in the coordinate frame co-rotating with the array with angular frequency $\omega$. In the dimensionless units used here $x$ and $y$ are scaled to a characteristic width $w_0 = 10$ μm, while propagation distance $z$ is scaled to the diffraction length $2\pi n w_0^2/\lambda \approx 1.14$ mm at the wavelength of $\lambda = 800$ nm and for background refractive index $n = 1.5$. Here dimensionless intensity $|\psi|^2 = 1$ corresponds to peak intensity of $I = n|\psi|^2/k^2 w_0 n_2$, where $n_2$ is the nonlinear refractive index of the material. Notice that rotating arrays can be fabricated using technology of fs-laser writing [7,25], applicable also to materials with defocusing nonlinearity. The function $\mathcal{R}(x,y) = \sum_n \mathcal{Q}(x-x_n, y)$ describes "line" SSH array composed from Gaussian waveguides $\mathcal{Q} = \exp[-(x^2+y^2)/\sigma^2]$ of width $\sigma = 0.5$. The potential depth is given by $p = \max[(2\pi w_0/\lambda)^2 \delta n/n] = 9$ that corresponds to $\delta n \approx 10^{-3}$. We vary the intracell $l_1$ and intercell $l_2$ distances between waveguides in dimers forming SSH array such that $l_1 + l_2 = 2d$, where $2d$ is the array constant. This affects coupling strength between waveguides inside each dimer and between dimers. We introduce dimerization parameter $\Delta = (l_1 - l_2)/2$ that quantifies the shift of the waveguides from the position where spacing $d$ between all waveguides in the array is equal.

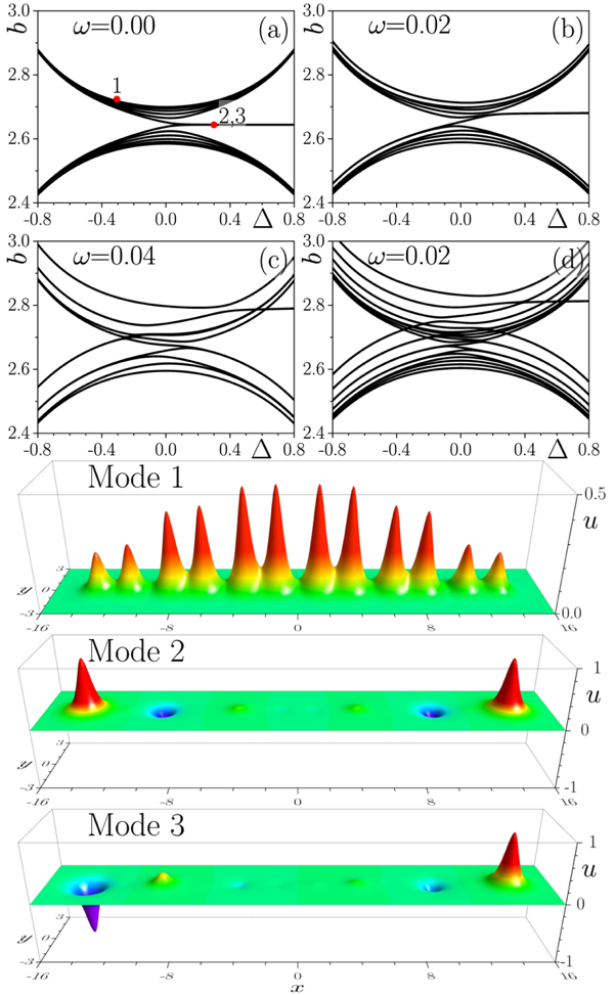

Fig. 1. Propagation constants $b$ of linear eigenmodes of the SSH array vs dimerization parameter $\Delta$ for different rotation frequencies $\omega = 0$ (a), $\omega = 0.02$ (b) and $\omega = 0.04$ (c) for the array with 12 waveguides ($N=6$ dimers) and $\omega = 0.02$ (d) for the array with 24 waveguides ($N=12$). Example of the bulk mode (mode 1) for nontopological non-rotating ($\omega = 0$) array with $\Delta = -0.3$ and examples of the edge modes with two in-phase peaks (mode 2) and two out-of-phase peaks (mode 3) in two outermost guides of the topological non-rotating array with $\Delta = 0.3$. Eigenmodes correspond to the red dots in (a). Here and below $p = 9$, $d = 2.4$, $\sigma = 0.5$.

We first omit the nonlinear term in (1) and search for linear modes as $\psi(x,y,z) = u(x,y)e^{ibz}$, where $b$ is the propagation constant, while the function $u$ describes the mode profile. Fig. 1(a)-(d) show representative dependencies of $b$ on the dimerization parameter $\Delta$ for various rotation frequencies $\omega$ and number of unit cells $N$ in truncated SSH array. For non-rotating array with $\omega = 0$ [Fig. 1(a)] a pair of topological edge states emerge in the topological gap at $\Delta > 0$, when the intercell waveguide spacing $l_2$ becomes smaller than the intracell one $l_1$ (see Supplement 1 for topological invariants). Examples of trivial bulk (mode 1) and topological modes (modes 2 and 3) are shown in Fig. 1. Mode 2 represents in-phase combination of the edge states at the opposite edges while mode 3 – their out-of-phase combination. Noteworthy, in topological modes the phase is inverted in neighboring unit cells, leading to specific staggered structure of tails. Array rotation modifies the spectrum of the system [Figs. 1(b), (c)], leading to band broadening and upward shift of the topological levels. Increasing the number of cells $N$ results in similar spectrum transformation since the impact of Coriolis term $\sim \omega$ on edge states becomes stronger in larger structure at fixed $\omega$ [Fig. 1(d)].

The dependence of $b$ on rotation frequency $\omega$ for $\Delta = 0.3$ and $\Delta = -0.3$ is shown in Figs. 2(a), (b). Blue dots indicate states localized at the edge of the array. For topological array with $\Delta = 0.3$ rotation shifts the edge levels upwards and, for large $\omega$, moves them out of the topological gap. In the rotating array, the eigenmode $u$ becomes complex and examples of its real and imaginary parts for different $\omega$ are illustrated in two top rows of Fig. 2. We show profile in the $x \geq 0$ domain only. Due to the term with $\omega$, the equation (1) is not invariant with respect to the transformation $y \to -y$, hence the modes are asymmetric with respect to the $y$-axis for $\omega \neq 0$. One can see from the figure that the mode from the gap retains the structure of tails representative for topological states, while in the semi-infinite gap the edge mode acquires in-phase tail. Trivial non-rotating array ($\omega = 0$ and $\Delta = -0.3$) does not support edge states, but in the presence of rotation they emerge in the semi-infinite gap [Fig. 2(b)]. With increase in $\omega$ the centrifugal effect causes the edge state to become more and more localized in the outermost waveguides (see modes 4 and 5). Although the mode 6 in trivial array appears in the gap, it is not true edge state, since in this mode the intensity in the outermost edge waveguide is always lower than intensity in the next-to-interface waveguide.

We now consider the formation of edge solitons in the presence of the *defocusing* nonlinearity in (1). They can be found in the form $\psi(x,y,z) = u(x,y)e^{ibz}$, where $b$ now is an independent parameter determining soliton shape and power $U = \iint |\psi|^2 dxdy$. Figure 3 illustrates nonlinear families for different array parameters. Gray dashed lines in Fig. 3 correspond to the propagation constants of linear bulk modes. We trace nonlinear families (and their "continuation" in other gaps of the spectrum) emerging from linear edge states: on this reason, when propagation constant of soliton approaches that of linear edge mode, its power $U$ vanishes. In Fig. 3(a) at $\Delta = 0.3$, $\omega = 0.02$ the edge soliton (see state 1) bifurcates from topological edge state and on this reason, it inherits a specific topological structure of tails. When $b$ shifts into the allowed band, coupling with bulk modes occurs and the soliton acquires a long tail in the array. This coupling results in the emergence of multiple $U(b)$ branches, two of which, with lowest powers, are shown in the gap below topological gap (see states 1 and 2 in Fig 1S from Supplement 1). Notice that these states feature different structure of tails in comparison with state 1 – the field changes its sign in neighboring waveguides, rather than in each second waveguide, as it occurs in topological state 1.

For higher rotation frequency $\omega=0.06$ the propagation constant of the linear edge state shifts into semi-infinite gap, while bulk band splits into a set of discrete separated levels shown by the dashed lines [Fig. 2(a)]. Now thresholdless edge soliton bifurcates from edge state with in-phase tail in semi-infinite gap (see state 4 in Supplement 1). One can see that complexity of the structure of soliton families increases at higher rotation frequencies $\omega$. In Fig. 3(b) we have omitted some hybrid families representing combinations of the edge and bulk states, since we are only interested in the simplest solitons localized at the edge of the array. One can see from the middle row of Fig. 3(b) that soliton branch 2 has the structure of tails typical for topological edge states, while edge solitons from the other branches feature different profiles, reminiscent to profiles of conventional gap solitons in 1D arrays (see states 3 and 4 in Supplement 1). Thus, one may conclude that defocusing nonlinearity can restore topological structure of the mode even if array rotation pushes it outside topological gap.

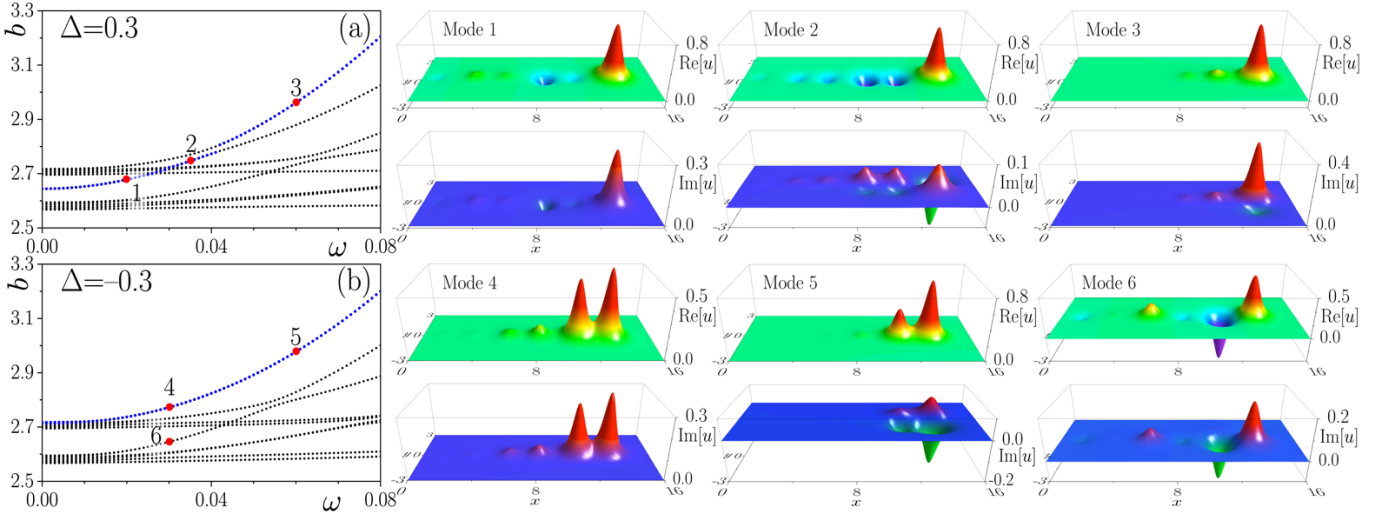

Fig. 2. Propagation constants b of the linear modes of the rotating SSH array vs rotation frequency $\omega$ for different dimerization parameters $\Delta=0.3$ (a), and $\Delta=-0.3$ (b). Branches shown with blue dots are associated with modes localized at the edge of the array. Examples of the real and imaginary parts of the modal fields at $\Delta=0.3$ (two top rows) and $\Delta=-0.3$ (two bottom rows) corresponding to the red dots in panels (a) and (b), respectively. Different color scales are used for the real and imaginary parts of $u$.

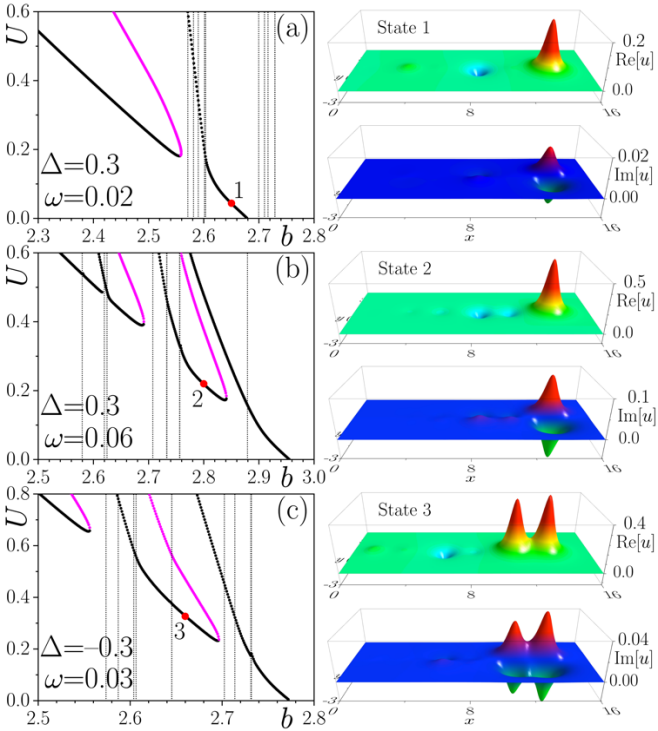

Fig. 3. Families of solitons bifurcating from linear edge modes in rotating arrays with $\Delta=0.3$, $\omega=0.02$ (a), $\Delta=0.3$, $\omega=0.06$ (b) and $\Delta=-0.3$, $\omega=0.03$ (c). Black families are stable, while magenta ones are unstable. Vertical dashed lines indicate propagation constants of the bulk states. Examples of the real and imaginary parts (right column) of the field for solitons corresponding to the red dots in panels (a)-(c). These nonlinear states feature structure of tails similar to that of topological edge states.

Furthermore, array rotation leads to the appearance of the edge states in a semi-infinite gap even in non-topological SSH array. Soliton family bifurcating from such nontopological edge states (see state 6 in Supplement 1) is presented in Fig. 3(c). Various branches representing "continuation" of the above family in other $b$ domains were encountered, among which is the branch 3, where solitons surprisingly acquire tails reminiscent of those for topological modes.

In Fig. 3 black branches correspond to stable solitons, while magenta branches to unstable ones. Stability was analyzed by modeling propagation in Eq. (1) of perturbed solitons $\psi|_{z=0} = u(x,y)(1+\delta_{\text{re}}+i\delta_{\text{im}})$, where $\delta_{\text{re}}+i\delta_{\text{im}}$ is a small noise, whose amplitude is uniformly distributed within the segment $[-0.05,+0.05]$. Figure 4(a) illustrates stable propagation of the edge solitons (including previously discussed states with topological structure), while Fig. 4(b) shows an example of evolution of the unstable state exhibiting nearly periodic amplitude oscillations due to instability.

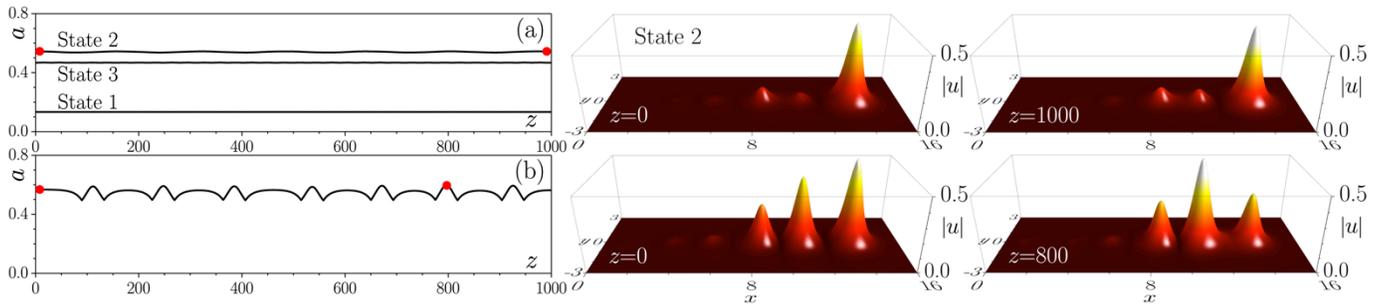

Fig 4. Propagation dynamics of perturbed stable topological (top row) and unstable (bottom row) edge solitons. Peak amplitude $a$ of (a) stable nonlinear states 1, 2, and 3 and (b) of unstable nonlinear state from Fig. 3(a) with $b=2.5$ (see state 2 in Supplement 1) versus distance are shown in the left column. Field modulus distributions for nonlinear stable state 2 (top row) and unstable state (bottom row) at different distances $z$ are shown in the middle and right columns. Profiles correspond to the red dots in panels (a), (b).

In closing, we have shown that rotation of topological SSH array considerably changes the structure of its spectrum leading to transition from modes of topological origin into modes that are localized due to array rotation. Defocusing nonlinearity enables formation of stable edge solitons, whose internal structure strongly depends on soliton power and rotation frequency.

**Funding:** Russian Science Foundation (grant 21-12-00096); Grants No. CEX2019-000910-S and No. PGC2018-097035-B-I00 funded by MCIN/AEI/10.13039/501100011033/FEDER, Fundació Cellex, Fundació Mir-Puig, and Generalitat de Catalunya (CERCA) and AGUR (2021 SGR 01448).

**Disclosures:** The authors declare no conflicts of interest.

**Data availability.** Data underlying the results presented in this paper are not publicly available at this time but may be obtained from the authors upon reasonable request.

**Supplemental document.** See Supplement 1 for supporting content.